\documentclass[reprint,english,twocolumn,notitlepage]{revtex4-1}
\usepackage[T1]{fontenc}  \usepackage[latin1]{inputenc}
\usepackage{babel} \usepackage{txfonts}
\usepackage{epsfig,epsf,psfrag} 
\usepackage{graphicx} 
\usepackage{subfigure}
\usepackage{pslatex}
\usepackage{epic,eepic} 
\usepackage{color,pstcol}
\usepackage{pstricks} 
\usepackage{times}

 \vspace{-10.0cm}  
 
\begin{document} \title{Optimal quantum refrigeration in strained graphene} 
  \author{Arjun Mani}
 \author{Colin Benjamin} \email{colin.nano@gmail.com}\affiliation{School of Physical Sciences, National Institute of Science Education \& Research, HBNI, Jatni-752050,\ India }
\begin{abstract}
 Refrigerators soak up heat energy from a low-temperature region and dump it into a higher temperature region using external work done on the system. Refrigerators are useful in cooling down a system to very low temperatures. In this letter, we show that a monolayer of strained graphene can be used in designing a quantum refrigerator with an excellent coefficient of performance and large cooling power. The operating point at which the strained graphene device works best as a quantum refrigerator is derived and the effects of strain and temperature on cooling are discussed. 
 \end{abstract}
\maketitle
\section{Introduction}
The efficacy of refrigerators at the nanoscale has been made more than obvious in the past half decade \cite{giu}. From being useful in schemes for removal of excess heat in nanosystems to achieving low temperatures or in designing quantum computers they have been one of the most productive areas of research\cite{tan}. {In these nanoscales, refrigerators which utilize quantum properties like wave nature of electrons are particularly useful. Since as systems size decrease, it will invariably lead to mesoscopic lengths where the wave nature of electrons is apparent\cite{heikkila}. In this context, we study a quantum refrigerator(QR) designed with a strained graphene layer, wherein the quantum property of wave nature of electrons is much more apparent than say in 2DEG's}. Even though, the thermoelectric figure of merit $ZT$ of graphene is very small around $0.01-0.1$\cite{yuri}, which is much smaller than some of the most efficient thermoelectric materials, e.g., Bi$_2$Te$_3$, see Refs.~\cite{kim,yuri}. The reason behind this small $ZT$ factor is its large thermal conductance and absence of any band gap. In some recent theoretical works\cite{tran,sadeghi,pohao,anno}, this factor $ZT$ has been predicted to rise to moderate value of around $2.5-3$. This improvement in the thermoelectric figure of merit is attributed to doping graphene with isotopes\cite{tran} or nanopores\cite{sadeghi,pohao} or disorder\cite{anno,tran} or by nano-patterning the graphene surface\cite{kim}. A large $ZT$ factor is required to generate high coefficient of performance (COP) in any refrigerator. A large COP means the refrigerator can use the electrical power to absorb heat energy from the cooler terminal more efficiently. Instead of doping as done in related works, we consider straining the graphene layer in order to generate this high COP. In a previous work of ours\cite{arjun}, we have shown the potential use of a strained graphene sample to operate as a quantum heat engine. In this manuscript, we concentrate only on the refrigeration aspect of a strained graphene monolayer.  

Another important aspect of this proposed strained graphene quantum refrigerator is that it operates in the steady state transport regime. Quantum refrigerator's (QR's) which work in the cyclic transport regime are also a major avenue of research. Examples of cyclic quantum refrigeration in literature can be seen in Refs.~\cite{lin, feng, he}. Cyclic QR's are of two types- 1) reversible, 2) irreversible. Advantage of cyclic reversible quantum refrigeration over both steady state quantum refrigeration and cyclic irreversible quantum refrigeration is that cyclic reversible  QR's, such as Carnot refrigerators and Otto refrigerators, are independent of the property of the working substance\cite{he}, i.e., the material characteristics, while all steady state QR's and cyclic irreversible QR's are dependent on the working substance. Cyclic QR's do have some disadvantages too. Cyclic reversible QR's like Carnot or Otto refrigerators are based on reversible processes, thus to complete one full cycle they take infinite time, thereby, reducing the practical application of these refrigerators. Cyclic irreversible QR's are dependent on irreversible processes which take much less time to complete one full cycle, however, the efficiency of these refrigerators is much reduced from the Carnot limit due to the dissipation within the system. On the other hand, examples of steady state QR's can be seen in Refs.~\cite{ya, wang, edward, koslof}. Steady state QR's absorb heat from the cooler terminal by moving the microscopic particles like electrons, phonons or photons, rather than moving any microscopic part of the system (like nano cyclic refrigerators). Since, steady state QR's do not depend on the movement of the microscopic body part of the system, they can be much smaller in size than the cyclic QR's. {In both steady state as well as cyclic QR's the efficiency can reach the Carnot limit but always in the limit of  zero cooling power. This happens due to the fact that cooling power and efficiency  depend on the electrical, thermal conductances and Peltier coefficient in such a way that if one tries to increase the cooling power to its maximum value then the efficiency reduces to a minimum and vice versa.}

The rest of the paper is arranged as follows. In section 2 we describe the theory needed to calculate the thermoelectric figure of merit $ZT$, Onsager coefficients, maximum COP and cooling power. The operating regime where our system works as a quantum refrigerator is also explained. Next in section 3 we describe our model of monolayer strained graphene sample and using the boundary conditions the transmission function for electrons between the reservoirs is derived. In section 4, we describe the results for our model and explained the reason behind getting large COP and cooling power observed in our system. Next in section 5 we discuss the experimental realization of our model. Finally, in section 6 we conclude our paper with a conclusion and a comparison with the results observed in other related proposals. 
\section{Theory of the quantum refrigerator}
\begin{figure}
  \centering {\includegraphics[width=0.48\textwidth]{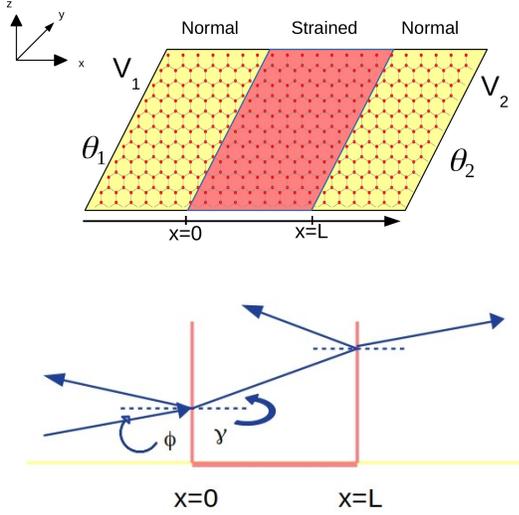}}
  \vskip -1.50in
\caption{Top: Monolayer graphene with uniaxial strain applied in the x direction. The middle portion is strained region while the two side portions are normal graphene regions. Voltages $V_1$ and $V_2$ are applied to the two terminals which are at temperatures $\theta_1$ (left side) and $\theta_2$ (right side) respectively. Bottom: An electron is incident on the interface between normal graphene and strained graphene with incident angle $\phi$ and refracted to strained region with refraction angle $\theta$.}
\label{fig1}
\end{figure}

To design an efficient  QR, apart from the fact that we are in length scales wherein the electron wave nature is apparent, we need a large Peltier coefficient along with a large Seebeck coefficient. Large Peltier and Seebeck coefficients are required to increase the $ZT$ factor and a large $ZT$ factor will engender a large COP. However, having large Seebeck and Peltier coefficients is a double edged sword, it reduces the cooling power of the system, i.e., the amount of heat energy which can be absorbed from the colder terminal. The cooling power only increases when the thermal and electrical conductances of the system increase with Seebeck and Peltier coefficients simultaneously decrease. To optimize these quantities in such a way that we get a large COP along with large cooling power is quite difficult, because all these parameters are inter-related. If we want to increase or decrease any one of these parameters, other parameters too are affected. So, it is one of the outstanding challenges in quantum thermoelectrics to design an efficient QR which would optimize these parameters effectively such that we get large COP and large cooling power. The difference between a quantum heat engine and QR is that in a quantum heat engine one always needs a small thermal conductance to get a large efficiency at maximum output power, while for quantum refrigeration one needs a large thermal conductance for large cooling power. Of course here we are talking about electronic contribution to the thermal conductance to be large, phonon contribution to the thermal conductance has to be small otherwise it will decrease both the COP and the cooling power. To calculate the COP and cooling powers and to design a QR, first we need to calculate the themoelectric properties of our strained graphene device, i.e., the Seebeck and Peltier coefficients alongwith the electrical and thermal conductances. In linear transport regime the electrical and heat currents are related to the electric and thermal biases via the Onsager coefficients, which are written as-\cite{ramshetti, benetti, sothmann}-
\begin{equation}\label{current}
\left(\begin{array}{c} J^e \\ J^{Q}\end{array}\right)=\left(\begin{array}{cc} L^{11} & L^{12}\\ L^{21} & L^{22} \end{array}\right) \left(\begin{array}{c} \mathcal{V}\\ \Delta \theta \end{array}\right),
\end{equation}
where $J^e$ and $J^Q$ define the electric and heat currents respectively, $L_{ij}$ with $i,j$ $\in$ $1,2$ denotes the Onsager coefficients, {$\mathcal{V}$ and $\Delta\theta$ are the potential bias and temperature bias applied to the system respectively}. The Seebeck coefficient is defined as the voltage difference generated across the system due to an unit temperature difference applied, while Peltier coefficient is defined as the ratio of the heat current transmitting through the junctions to the electrical current passing through that junction. They are given as follows-
\begin{equation}\label{Seeback}
S=-\frac{L^{12}}{L^{11}}, \quad\text{and}\qquad P=\frac{L^{21}}{L^{11}}.
\end{equation}
The Onsager co-efficient matrix written in Eq.~(\ref{current}) linking electric and heat currents to the temperature difference ($\Delta\theta$) and applied voltage bias ($\mathcal{V}$) thus can be rewritten as \cite{ramshetti,mazamutto}-
\begin{equation}\label{onsager}
\left(\begin{array}{cc} L^{11} & L^{12}\\ L^{21} & L^{22} \end{array}\right) =\left(\begin{array}{cc} \mathcal{L}^{0} & \mathcal{L}^{1}/e\theta\\ \mathcal{L}^{1}/e & \mathcal{L}^{2}/e^2\theta \end{array}\right),
\end{equation}
wherein,
\begin{eqnarray}\label{con}
\mathcal{L}^{\alpha}=G_0\int_{-\pi/2}^{\pi/2}\!\!\!d\phi \cos\phi \int_{-\infty}^{\infty}\!\!\!d\epsilon(-\frac{\partial f}{\partial \epsilon})\frac{|\epsilon|}{\hbar v_f}(\epsilon-\mu)^\alpha T(\epsilon,\phi),
\end{eqnarray}
here $G_0=(e^2/\hbar)(W/\pi^2)$, $\mathcal{L}^0=G$ defines the conductance of system with width of sample-$W$\cite{dolphas}, $\epsilon$-the electron energy, $f$-the Fermi-Dirac distribution function, $\phi$ defines the angle of incidence for electrons, $T(\epsilon,\phi)$ is the electronic transmission probability through strained graphene layer and $\mu$- the Fermi energy. Eq.~(\ref{con}) is for a sheet of monolayer graphene (2D system) with the transmission function($T$) defined between the two terminals, see also Fig.~1. Once we know the transmission probability-$T(\epsilon,\phi)$, we can calculate the Onsager coefficient's- $L^{ij}$'s as shown in Eq.~(\ref{current}). After calculating the Onsager coefficients, other quantities such as cooling power and COP can be calculated as follows. The cooling power \cite{benetti}, is defined as -
\begin{eqnarray}\label{power}
{J^Q}=(L^{21}\mathcal{V}+L^{22}\Delta \theta) 
\end{eqnarray}
Our strained graphene mono-layer device is shown in Fig.~\ref{fig1}. It works as a quantum refrigerator(QR) only when a net heat current is flowing from the cooler to hotter terminal making the colder terminal still more cooler, i.e., when it flows against the applied temperature bias $\Delta \theta=\theta_1-\theta_2$ ({$\theta_i$ is the temperature at contact $i$}) and the net electrical current flows from higher potential bias to the lower bias, i.e., the external work is done on the system. Thus, if a temperature bias $\Delta \theta$ is applied at the left terminal ($\theta_1>\theta_2$) and potential bias $\mathcal{V}$ is applied at the right terminal ($V_2>V_1$), then both heat current and electrical current flow from right to left ($J^e<0, J^Q<0$ considering $+x$ direction as the positive direction) for this system to work as a QR. The efficiency of a QR, i.e., how good it is in converting a stream of charged particles into carrying heat energy is called coefficient of performance (COP)\cite{edward,koslof}. COP of a quantum refrigerator is defined as the ratio of heat current absorbed from the hot reservoir to electrical power-$\mathcal{P}$ applied on the system, such that-
\begin{eqnarray}
\eta^r=\frac{J^Q}{\mathcal{P}},
\end{eqnarray} 
where $\mathcal{P}$, is the electrical work or power done on the system, is defined as-
\begin{eqnarray}
\mathcal{P}=J^e\mathcal{V}=(L^{11}\mathcal{V}+L^{12}\Delta \theta)\mathcal{V}.
\end{eqnarray}
The COP-$\eta^r$ is maximum when $\frac{\partial \eta^r}{\partial \mathcal{V}}=0$ and considering $J^Q<0$ and $\mathcal{P}<0$, we have-
\begin{eqnarray}
\mathcal{V}&=&\frac{L^{22}}{L^{21}}(-1-\sqrt{\frac{L^{11}L^{22}-L^{12}L^{21}}{L^{11}L^{22}}})\Delta \theta.
\end{eqnarray} 
The COP becomes maximum when the above relation Eq.~(8) between the potential bias and thermal bias holds and thus the maximum COP is when-
\begin{eqnarray}
\quad\eta^r_{max}&=&\frac{\eta^r_c}{x}\frac{\sqrt{ZT+1}-1}{\sqrt{ZT+1}+1},\\
\text{with, }\quad J^Q(\eta^r_{max})&=&\sqrt{\frac{L^{22}(L^{11}L^{22}-L^{12}L^{21})}{L^{11}}}\Delta \theta,
\end{eqnarray} 
wherein $\eta^r_c=\frac{T}{\Delta \theta}$ is defined as the Carnot efficiency of an ideal refrigerator and $J^Q(\eta^r_{max})$ is the cooling power when COP is maximum. The thermoelectric figure of merit $ZT$ is defined as-
\begin{eqnarray}\label{ZT}
ZT=\frac{G S^{2}}{\kappa} \theta,
\end{eqnarray}
while thermal conductivity is-
\begin{eqnarray}\label{kappa}
\kappa=\frac{L^{11}L^{22}-L^{12}L^{21}}{L^{11}}.
\end{eqnarray}
  In multi-terminal systems with broken time reversal(TR) symmetry, the upper bound on the refrigerator efficiency $\eta^r_{max}$ decreases from $\eta^r_c$ as the asymmetric parameter $\mathcal{\chi}=\theta L^{12}/L^{21}$ deviates from one \cite{brandner, bran}. In multi terminal systems with TR symmetry preserved and for all two terminal systems irrespective of whether TR symmetry is broken or not, the asymmetric parameter $x$ is unity, and upper bound on the corresponding maximum efficiency-$\eta^r_{max}$ equals $\eta^r_c$. This is the advantage of multi-terminal systems which preserve TR symmetry and for any two-terminal system, that they can work as highly efficient QR's with almost Carnot efficiency. However, for multi-terminal systems with broken TR symmetry working as QR, the upper bound is always less than $\eta^r_c$(the Carnot limit). To work as a QR, the relation between potential bias $\mathcal{V}$ and thermal bias $\Delta \theta$ has to be such that the electrical current $J^e<0$ and the heat current $J^Q<0$, i.e., both electrical and heat current flow from the cooler to the hotter region. Solving these two equations $J^e<0$ and $J^Q<0$ we get the operating regime for QR's as $\mathcal{V}<-(\kappa/(GP)-S)\Delta\theta$. This defines the parametric space in which our graphene device acts as a QR. In the next section, we give a detailed description of our model. 
\section{Model}
\subsection{Hamiltonian}
Graphene is a semi-metal with zero band gap. It is a carbon allotrope with carbon atoms arranged in a single layer of honey comb lattice with inter penetrating triangular sublattices.  A sublattice is a non-empty subset of a lattice, thus, a crystal can have several independent sublattices. Due to the presence of these two independent sublattices, graphene does have two non-equivalent set of local minima (Dirac point) in its momentum space at the edge of the first Brillouin zone. These two sets of local minima (Dirac points) are called as $K$ and $K'$ valleys. Dirac points are those points in the momentum space where conduction band meets with the valence band.

An uniaxial strain is introduced in our model of mono-layer graphene sheet lying in the $xy$ plane via stretching or compressing the region between $x=0$ and $x=L$ as shown in Fig.~1. The region to the left and right of this strained region is normal graphene. To design our model as a nano refrigerator we apply thermal bias $\Delta \theta$ at the left contact and a potential bias $\mathcal{V}$ at the right contact. At steady state, electric current $J^e$ and heat currents $J^Q$ flow between the reservoirs. In the strained region, electrons gets refracted away from the normal (perpendicular to the interface) in one valley (say $K$) and refracted towards the normal in the other valley ($K'$). The transmission of electrons gets shifted in the two valley as a function of incident angle in two opposite directions. However, the total transmission which is sum over all the incident angles remains same for both the valleys $K$ and $K'$. The Hamiltonian for the system for $K$ and $K'$ valley is-
\begin{equation}\label{ham}
\mathcal{H}_{K}=\hbar v_f\sigma(k-s'), \quad \mathcal{H}_{K'}=-\hbar v_f\sigma^*(k+s').
\end{equation}
Considering Landau gauge, one can replace the strain with pseudo magnetic vector potential $A=(0, \pm A_y)$, where `+' and `-' signs denote $K$ and $K'$ valleys respectively\cite{castro}. {The parameter $s'$ in Eq.~(13) is related to the strain $s$ by the relation $s=\hbar v_f s'$, and $s$ can be defined as the perturbation to the nearest neighbor hopping amplitudes, $\delta t$. Throughout our manuscript we have defined strain as $s$ in units of $meV$, if we divide the strain $s$ by the nearest neighbour hopping amplitude $t$ (which is a constant and equal to $2.7 meV$) then we will get the strain as a dimensionless quantity. Thus, strain-$s=\hbar v_f s'=eA_y[\Theta(x)-\Theta(x-L)]$, with $v_f$-the Fermi velocity, $\sigma=(\sigma_x, \sigma_y)$ are Pauli matrices operating on the sub-lattices A and B with $\sigma^*$ being complex conjugation, $\Theta$-the Heaviside step function and $k(=\{k_x, k_y\})$-the 2D wave vector.} From Hamiltonian, Eq.~(\ref{ham}), we get the wave equation for $K$ valley-
\begin{eqnarray}\label{wav}
\hbar v_f(-i\partial x-\partial y-i s)\psi_B=E\psi_A,\nonumber\\
\hbar v_f(-i\partial x+\partial y+i s)\psi_A=E\psi_B,
\end{eqnarray}
where $\psi_A$ and $\psi_B$ are the wave functions at $A$ and $B$ sublattices respectively. Using Eq.~(\ref{wav}), one can calculate the transmission probability for ballistic transport in a monolayer of strained graphene sample, which we show in the next subsection. 
\subsection{Wave functions and boundary conditions}
In Fig.~1(bottom) an electron is incident from the left side of the interface between two regions (region I and region II) with energy $\epsilon$, then either it can reflect back to region I or it can transmit into region II depending on its energy and angle of incidence. We define three regions I, II and III as normal graphene ($x<0$), strained graphene ($0<x<L$) and normal graphene ($x>L$) respectively. The wave functions of electrons in the three regions for A and B sublattices in $K$ valley are as follows-\\ 
For $x<0$-
\begin{eqnarray}\label{a}
\left[\begin{array}{c}\psi_A^1(x,y)\\\psi_B^1(x,y)\end{array}\right]=\left[\begin{array}{c} (e^{ik_xx}+r e^{-ik_xx}) \\(e^{ik_xx+i\phi}-r e^{-ik_xx-i\phi})\end{array}\right]e^{ik_yy},
\end{eqnarray}
in strained graphene layer $0<x<L$-
\begin{eqnarray}\label{b}
\left[\begin{array}{c}\psi_A^2(x,y)\\\psi_B^2(x,y)\end{array}\right]=\left[\begin{array}{c} (a e^{iq_xx}+b e^{-iq_xx}) \\ (a e^{iq_xx+i\theta}-b e^{-iq_xx-i\theta}) \end{array}\right]e^{ik_yy},
\end{eqnarray}
and for $x>L$-
\begin{eqnarray}\label{c}
\left[\begin{array}{c} \psi_A^3(x,y)\\\psi_B^3(x,y)\end{array}\right]=\left[\begin{array}{c} t e^{ik_xx} \\t e^{ik_xx+i\phi} \end{array}\right]e^{ik_yy},
\end{eqnarray}
where $k_x=(\epsilon/\hbar v_f)\cos\phi$ and $k_y=(\epsilon/\hbar v_f)\sin\phi$ are the $x$ and $y$ components of momentum wave vector in normal graphene. In strained graphene $k_x$ is replaced with $q_x=\sqrt{(\epsilon/\hbar v_f)^2-(k_y-s)^2}=(\epsilon/\hbar v_f) \cos \gamma$ and $k_y-s=(\epsilon/\hbar v_f) \sin \gamma$, $\gamma$ being the refraction angle in the strained region as shown in Fig. 1(bottom) and also satisfies $\tan \theta=(k_y-s)/q_x$. Using wave functions- Eqs.~(\ref{a}-\ref{c}), and applying boundary conditions at $x=0$-
\begin{eqnarray} \label{ba}
\psi_B^2(x=0)=\psi_B^1(x=0), \quad
\psi_A^2(x=0)=\psi_A^1(x=0),
\end{eqnarray}
and at $x=L$- 
\begin{eqnarray} \label{bb}
\psi_A^2(x=L)=\psi_A^3(x=L), \quad \psi_B^2(x=L)=\psi_B^3(x=L).
\end{eqnarray}
Solving Eqs.~(\ref{ba}-\ref{bb}) we derive the electronic transmission probability for $K$ valley- 
\begin{eqnarray} \label{t}
T(\epsilon,\phi)=\frac{1}{\cos^2[q_xL]+\sin^2[q_xL](\frac{1-\sin[\gamma]\sin[\phi]}{\cos[\gamma]\cos[\phi]})^2}.
\end{eqnarray}
Similarly, one can derive the transmission function by solving the Hamiltonian for $K'$ valley as in Eq.~(\ref{ham}), using the same boundary conditions and substituting $\phi\rightarrow-\phi$, $s\rightarrow-s$. The total electronic transmission probability $T(\epsilon)$ is the sum of $K$ and $K'$ valley transmissions. It is apparent that although transmission $T(\epsilon,\phi)$ differs in $K$ and $K'$ valley\cite{can}, when integrated over $'\phi'$ this difference disappears. Thus, total transmission probability $T(\epsilon)$ is twice that of $K$ valley transmission.
   \begin{figure}
  \centering {\includegraphics[width=0.5\textwidth]{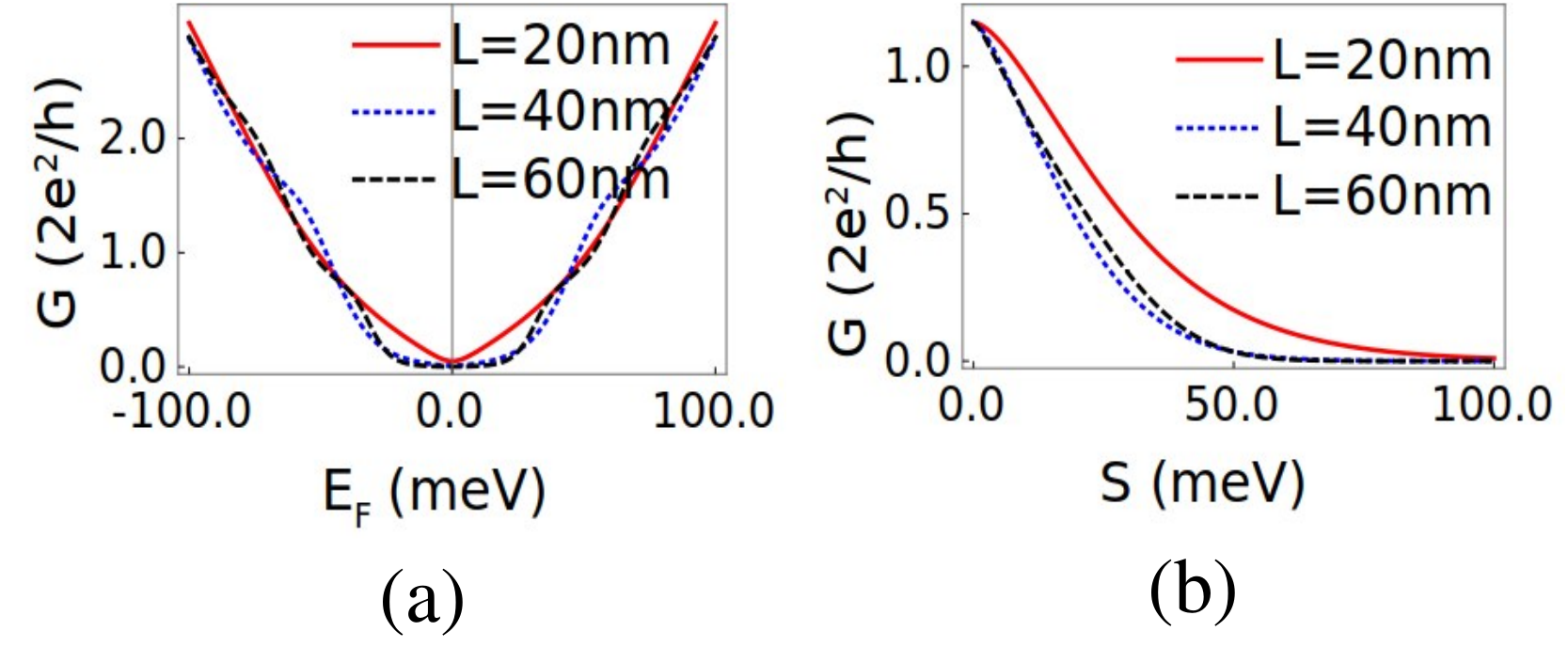}}
 \caption{(a) Conductance in units of $2e^2/h$ at temperature $\theta=30$K for various lengths of strained graphene layer with width $W=20$nm and strain $s=30$meV, (b) Conductance in units of $2e^2/h$ at temperature $\theta=30$K for various lengths of strained graphene layer with width $W=20$nm and Fermi energy $\mu=29.6$meV.}
\end{figure}
\section{Results and Discussion}
 To design an efficient quantum refrigerator(QR), we not only need large Seebeck coefficient but also a large Peltier coefficient. A large Peltier coefficient means that a stream of electrons or holes carrying current from one terminal to another not only transport charge but also heat energy along with them. Thus, decreasing the temperature of the colder terminal. In any  multi terminal system breaking time reversal symmetry implies an asymmetric Onsager matrix. However, for a two terminal system, Onsager matrix is always symmetric regardless of whether time reversal(TR) symmetry is broken or not\cite{datta}. This implies that the Seebeck and Peltier coefficient, for our system as shown in Fig.~1, are related by the relation $P=\theta S$ (due to the Onsager reciprocity relation $L_{21}=\theta L_{12}$). In Figs.~3 (a, b) we plot Peltier coefficient. The Seebeck coefficient can be inferred from these plots via $S=P/\theta$. Breaking TR symmetry leads to the upper bound of COP reducing from Carnot limit of efficiency, while in our system due to the conservation of TR symmetry the upper bound of COP can in principle reach the maximum Carnot limit. We first (in Fig.~2) plot the electrical conductances and see that it increases with increasing Fermi energy, see Fig.~2(a). We also see in Fig.~2(a), that increasing the length of the strained layer shifts the transport of incident electrons with low energy into the evanescent regime, and thus opens a conduction gap close to the Dirac point, while it remains almost unaffected at higher Fermi energies. Increasing strain can make the transmission of an electron at particular incident angle to be unity due to Klein effects\cite{kat}, however, the overall transmission summed over all the incident angle reduces with strain, see Fig.~2(b).  

The Peltier coefficient is the ratio of off-diagonal Onsager matrix element $L_{21}$ to the electrical conductance. Increasing strain reduces the electrical conductance (see Fig.~2(b)) which, in turn, increases the Peltier coefficient (see Fig.~3(a)). However, increasing Fermi energy increases the electrical conductance, this in turn, leads to an increase in the Peltier coefficient near the Dirac point. Near the Dirac point, there is an imbalance between electronic contribution and hole contribution to the Peltier coefficient, which increases the Peltier coefficient initially. However, as one goes away from the Dirac point this effect vanishes and the Peltier coefficient decreases gradually.
 \begin{figure}
  \centering {\includegraphics[width=0.5\textwidth]{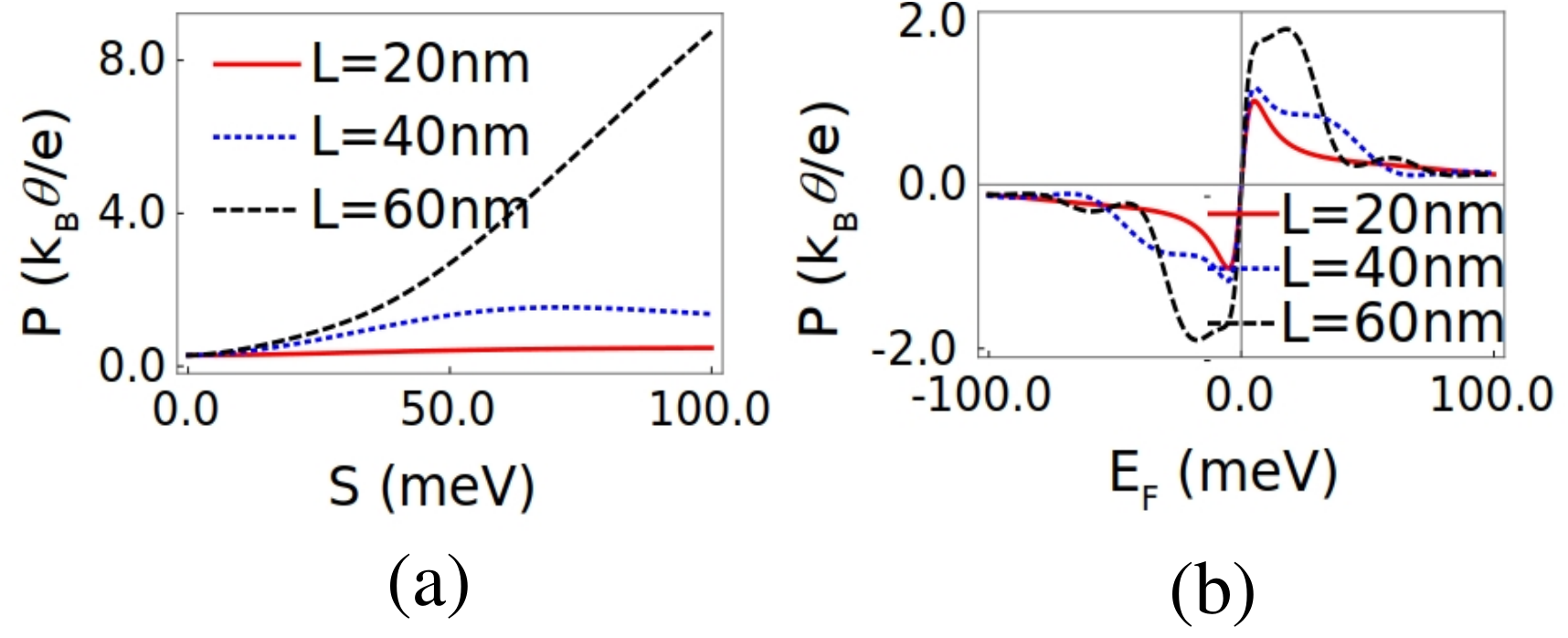}}
 \caption{(a) Peltier coefficient P in units of ($k_b\theta/e$) at temperature $\theta=30$K for various lengths of strained region with width $W=20$nm and Fermi energy $\mu=29.6$meV, (b) Peltier coefficient P in units of ($k_b\theta/e$) at temperature $\theta=30$K for various lengths of strained layer with width $W=20$nm and strain $s=30$meV.}
\end{figure}
This reduction of electrical conductance is almost independent of the length of the strained region. {In Refs.~\cite{mik, wu} it has been shown that strain can give rise to a resonant tunneling for some particular energy of incident electrons, thus acting like a quantum dot or quantum wire. However, this will not affect the performance of our device since we are taking an average over all the incidents angles. Resonant tunneling  occurs only for some particular energy and incident angle of the electron. If we take an average over all the incident angles then the overall transmission probability always reduces with increasing strain, and thus to vanishing resonance effects.} In Figs.~3(a) and (b), we see that increasing either strain or the length of the strained region makes electrical conduction slowly to shift to the evanescent regime, then electrical conduction reduces enabling Peltier coefficient to increase since it is inversely proportional to conductance. This means electrons carry large amount of heat energy along with them and thus we see a corresponding increase in the Peltier coefficient. As strain is increased more, the electrical transmission $T(\epsilon, \phi)$ shifts completely to the evanescent regime, and Peltier coefficient increases much more. This is the reason why, COP of our model is huge for larger strains and large lengths of the strained region, see Fig.~4(a), while the cooling power decreases, see Fig.~4(b). The positive Peltier coefficient for Fermi energies greater than the Dirac point energy observed in Fig.~3(b) is attributed to the increased electronic contribution while for Fermi energies below the Dirac point energy the negative Peltier coefficient is attributed to increasing hole contribution. The peaks in Peltier coefficient observed near the Dirac point for both positive as well as negative energies is due to the imperfect cancellation of electron and hole contributions to the Peltier coefficient, while for energies much higher or much lower than the Dirac point the electronic and hole contribution to the Peltier coefficient almost cancel leading to a vanishing Peltier coefficient.     
{
According to Eq.~(9) of our manuscript, coefficient of performance is proportional to the thermoelectric figure of merit $ZT$,which is the product of electrical conductivity $G$ and the square of Seebeck coefficient $S$, while inversely proportional to the thermal conductivity $\kappa$, i.e.,-
\begin{equation}
ZT=\frac{GS^2}{\kappa}\theta
\end{equation}
In the aforewritten equation, $\theta$ is the temperature. According to the Wiedemann-Franz law, for metals, electrical conductivity $G$ is proportional to the thermal conductivity $\kappa$ at finite temperature. Since, Graphene is a semi-metal, for its also $G$ is proportional to the $\kappa$. So, the ratio $G/\kappa$ is almost constant at finite temperature $T$. Thus, thermoelectric figure of merit $ZT$ solely depends on the square of the Seebeck coefficient. Since, graphene is a Dirac material, its Dirac cones shift in the opposite direction in two valleys upon applying strain creating a conduction gap close to the Dirac point, see Fig.~2(a). This conduction gap increases with increasing strain. Thus, the Seebeck (or Peltier) coefficient also increases with strain, since Seebeck (Peltier) coefficient is inversely proportional to the electrical conductivity $G$, see Eq.~(2). So, thermoelectric figure of merit $ZT$ as well as the coefficient of performance both increase with strain. However, this effect is not monotonous, if we keep on increasing strain then the ratio $G/\kappa$ is no more a constant and it reduces the $ZT$ as well as the coefficient of performance, see Fig.~4(a).}

In Fig.~5(a), we see that COP of our model is not temperature independent, the COP reduces as temperature decreases. On the other hand the cooling power is almost temperature independent, see Fig.~5(b). { The reason being that as we decrease temperature, the thermal conductance of the device remains almost unaffected by the strain, and thus the cooling power too, see Eqs.~(10, 12). However, the thermoelectric figure of merit of the device reduces with the temperature due to reduction in Peltier (Seebeck) coefficient and thus the COP reduces with temperature, see Eqs.~(9, 11).} In Fig.~6(a), we see that two peaks appear in the COP as function of Fermi energy. The first peak is present even at zero strain, that means strain is not a source of this peak. This peak appears close to the Dirac point and is due to the imbalance in the contribution of hole and electrons towards the Peltier coefficient as explained above. The second peak increases with increasing strain and vanishes for zero strain, this means strain is the sole reason behind the peak. The reason behind the appearance of the second peak is for the case of Peltier coefficient dominating over the electrical conductance, that's why the second peak increases with strain as Peltier coefficient increases. In Fig.~6(b), we see that the cooling power increases as a function of Fermi energy, but decreases with strain. So, we see that strain helps in increasing the COP but on the other hand it decreases the cooling power. Thus, we have to choose our parameters in such a way that we get the optimum values for COP and cooling power, i.e., the COP can still be large and the cooling power not so small. That is why we have defined a `Q' point for our quantum refrigerator where our device works at optimum values. `Q` point is the optimal operating range of our quantum refrigerator. Operating the refrigerator at the `Q' point, entails good COP as well as cooling power. The `Q' point of our refrigerator is at strain $s=30$ meV, with length of the strained region $L=60$ nm, temperature $\theta=30$ K, Fermi energy $\mu=29.6$ meV and width of graphene sample $W=20$ nm. For the `Q' point parameters, $COP=0.1 \eta_c^r$ and cooling power $J^Q=2 k_B^2\theta\Delta\theta/h$. However, our device exhibits a maximum $COP$ of $0.95\eta_c^r$ as seen in Fig.~7 but cooling power in this parametric regime is very small. Further, maximum cooling power possible in our system is $12k_B^2\Delta\theta/h$ as in Fig.~6(b) but COP at this parameter value is very small as shown in Fig.~6(a).  The discussion so far hasn't addressed the width of the sample. From Eq.~(4) of our manuscript we can see that the Onsager coefficients linearly depend on sample width $W$. However, since Seebeck, Peltier coefficients and figure of merit $ZT$  are ratios of these Onsager coefficients, all of them won't depend on the width $W$ of the system. Thus, coefficient of performance will remain unaffected due to the changes in width, however, the cooling power will increase linearly with the width of the system. We have chosen $W=20$ nm, which is a reasonable value for experimental applications.

{ Finally, we address the reasons for neglecting the phonon contribution in the Figures presented in this section.  The primary reason for this is because the thermal conductance in graphene is quite small (almost absent) at low temperatures $0-30 K$, see Figs.~2,3  in Ref.~\cite{mazamutto} and Fig.~5 of Ref.~\cite{xu}. Beyond $25-30 K$ range, the phonon contribution increases linearly with temperature, as shown in Refs.~\cite{mazamutto,zuev}. Thus, the phonon contribution to the thermal conductance can be neglected at the temperature range $20-30 K$ discussed in our work.}

 \begin{figure}
  \centering {\includegraphics[width=0.5\textwidth]{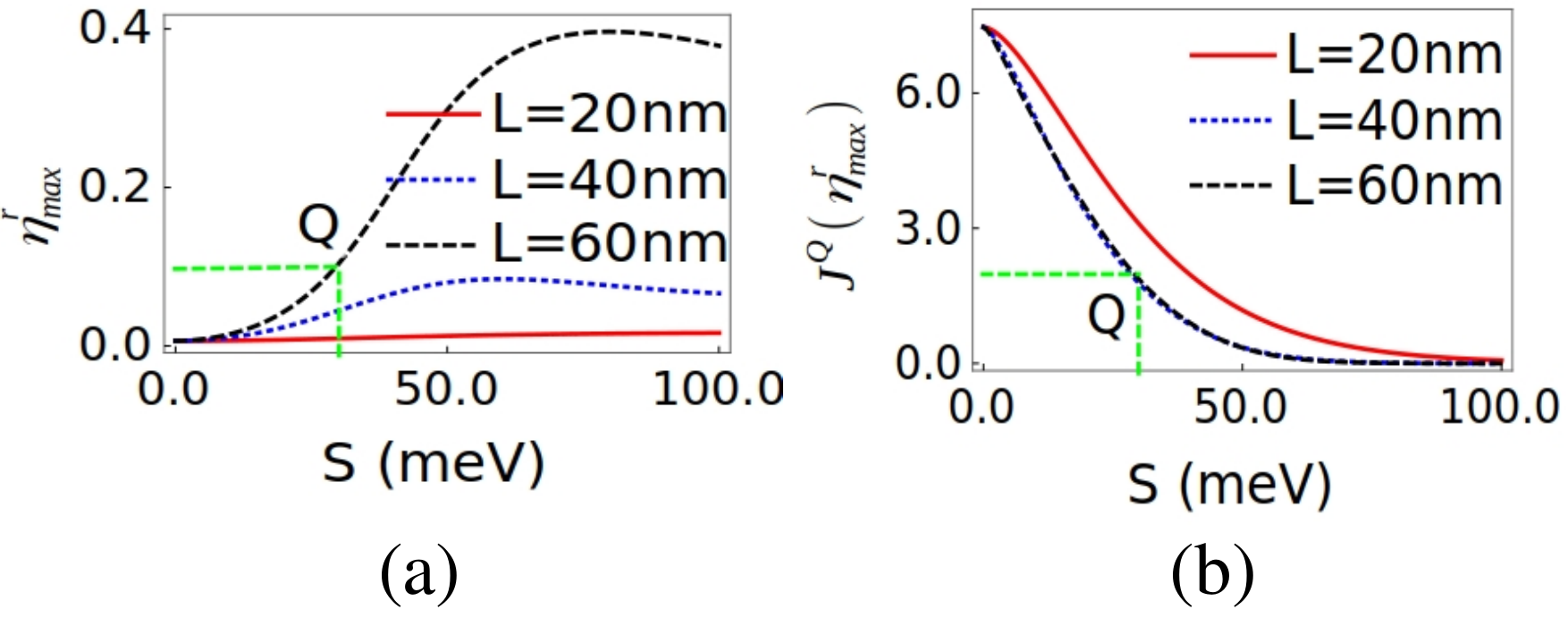}}
 \caption{(a) COP in units of $\eta_c^r$ at temperature $\theta=30$K for various lengths of strained layer with width $W=20$nm and Fermi energy $\mu=29.6$meV, (b) Cooling power in units of $(k_B^2\theta\Delta\theta)/h$ at $\theta=30 K$ for various lengths of the strained layer with width $W=20$nm and Fermi energy $\mu=29.6$meV.}
\end{figure}
 \begin{figure}
   \centering {\includegraphics[width=0.5\textwidth]{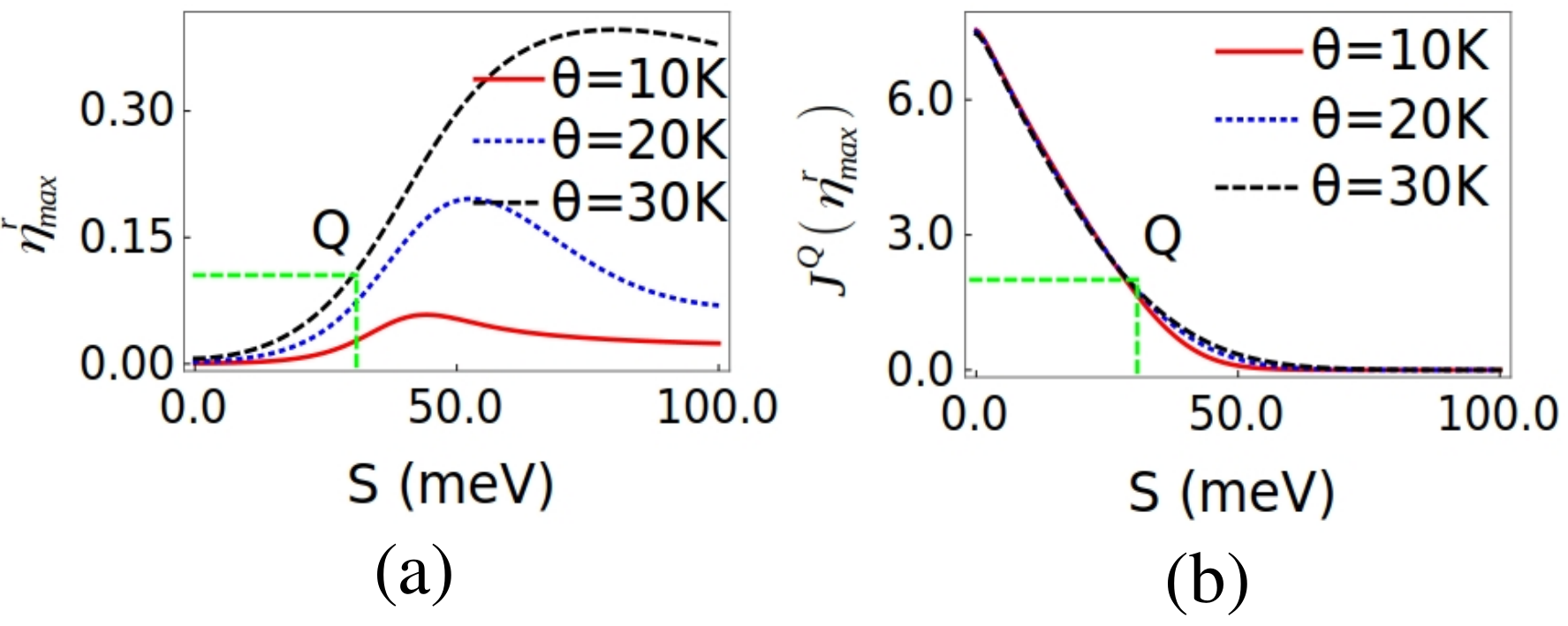}}
 \caption{(a) COP in units of $\eta_c^r$ at different temperatures with L=60 nm, width $W=20$ nm and Fermi energy $\mu=29.6$ meV, (b) Cooling power in units of $(k_B^2\theta\Delta\theta)/h$ for various temperatures with L=60 nm, width $W=20$ nm and Fermi energy $\mu=29.6$ meV.}
\end{figure}
 \begin{figure}
   \centering {\includegraphics[width=0.5\textwidth]{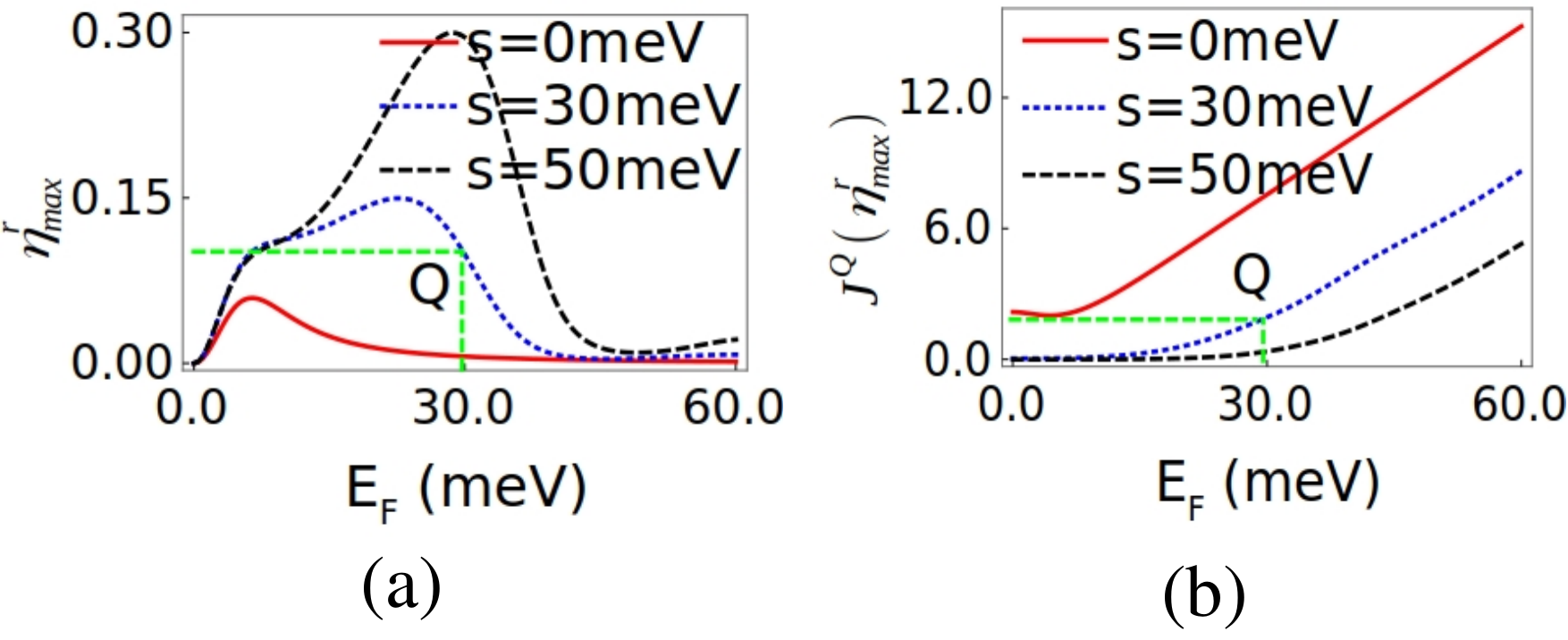}}
 \caption{(a) COP in units of $\eta_c^r$ at $\theta=30$K at different strains with $L=60$nm, width $W=20$nm, (b) Cooling power in units of $(k_B^2\theta\Delta\theta)/h$ at $\theta=30$K for various strains with $L=60$nm, width $W=20$nm.}
\end{figure}
\section{Experimental Realization}
Our proposal of a quantum refrigerator based on a strained monolayer graphene layer is experimentally realizable. Many theoretical, see Ref.~\cite{castro, nanoscale}, as well as many experimental works\cite{tmg, chhikara}, deal with uniaxial strain in monolayer graphene system. Thus, realizing strain in graphene system would not be much difficult. Also, the amount of strain used in our system is very small. One can apply a maximum $20\%$ strain ($540$ meV) without opening a band gap, while in our paper the optimum strain applied at `Q' point is $30$ meV ($1\%$ strain)\cite{liu,per-tight}. Further confining strain region only to the middle part has been attempted before too by stretching, compressing
or suspending the middle part of the graphene only without affecting the nonstrained region, see Ref.~\cite{balandin}. In Ref.~\cite{balandin}, only the middle part of the graphene sample is suspended across a wide trench in Si substrate, generating a finite strain but limited to the middle part only, see  Box. 1(a) (on page 573 of  Ref.~\cite{balandin}).

We have considered a sharp strain potential between the two normal regions for convenience only. One can also consider a small slope to the strain potential rather than a sharp strain potential. However, it has been shown that a small slope to the strain at the boundary between strain and normal regions also leads to the same amount of tunneling suppression for the electrons, see Refs.~\cite{castro, ade}. Recently, there have been other works, which show that strain can be reliably and easliy controlled in graphene, see Ref.~\cite{zhni}. Numerical values of all the other parameters used in our paper are also physically realizable and used in other works too, see Refs.~\cite{castro,dolphas}.    
\section{Conclusion and Perspective on related works}
In this manuscript a strained monolayer graphene sheet has been designed to work as a quantum refrigerator(QR). The maximum coefficient of performance of our QR is around $0.95 \eta_c^r$ (see Fig.~7). This large COP seen in our paper occurs at a strain $s=300$ meV, which is $11\%$ strain, much less than the maximum 20\% strain ($540$ meV) and thus does not open any band gap within our system. The maximum cooling power possible with our QR is $12(k_B^2\Delta\theta)/h$ (see Fig.~6(b)). However, we note that the maximum values of coefficient of performance and of cooling power seen in our QR do not correspond to same set of parameters, the optimum values of coefficient of performance as well as cooling power are $0.1 \eta_c^r$ and $2(k_B^2\theta)/h$ and these occur at identical parametric values, i.e.,  the`Q' point. Steady state quantum refrigeration as seen in this manuscript in strained graphene is a nascent topic. Not many works\cite{ya,wang,giazotto} have addressed the topic of steady state quantum refrigeration. In the following we briefly address those works mentioning their advantages and disadvantages vis-a-vis our work.
 \begin{figure}
  \centering {\includegraphics[width=0.45\textwidth]{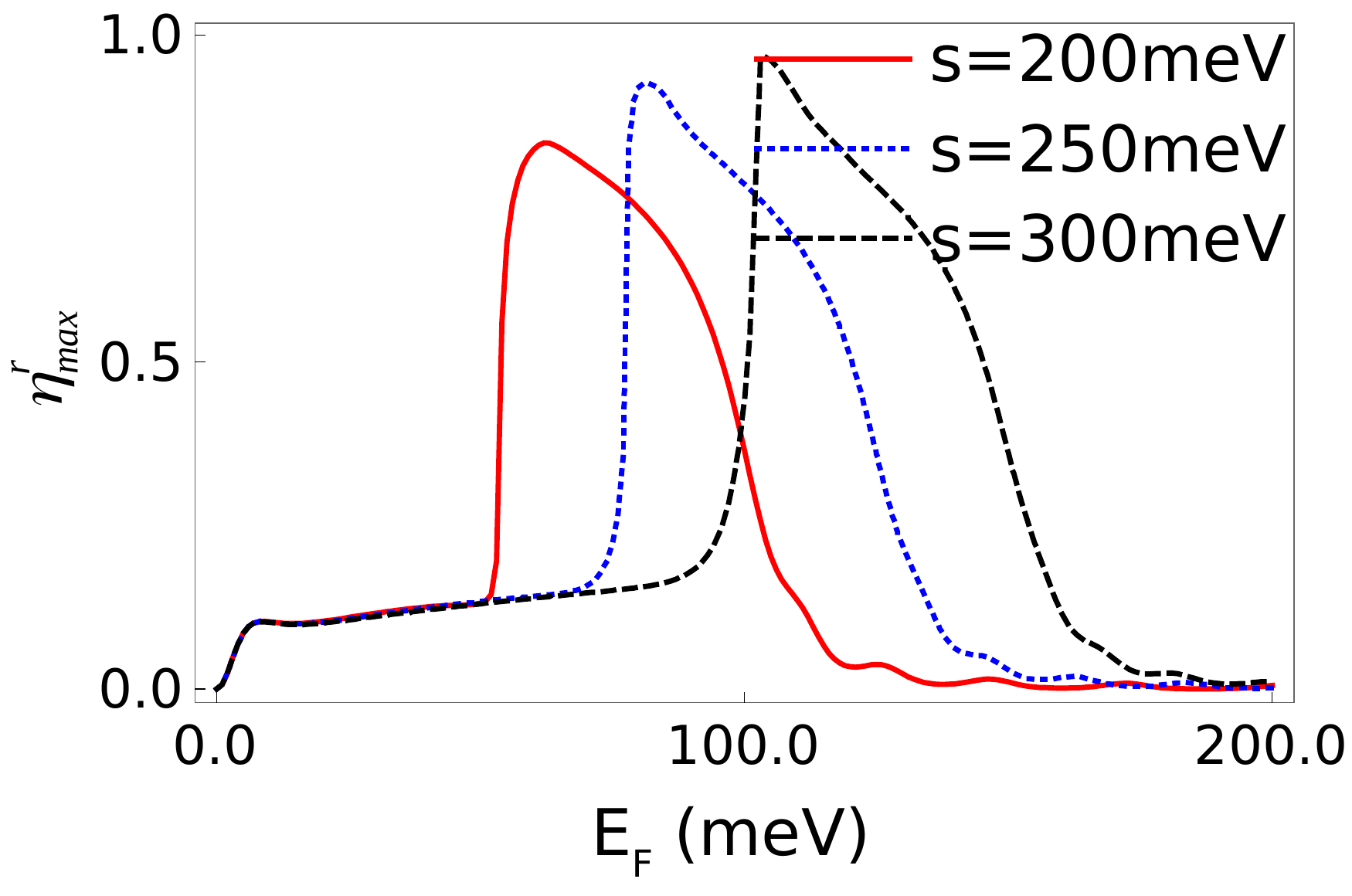}}
 \caption{COP in units of $\eta_c^r$ at $\theta=30$K at different strains with $L=60$nm, width $W=20$nm.}
\end{figure}
Ref.~\cite{ya} discusses a three terminal quantum dot refrigerator, wherein the maximum COP and cooling power are $0.4\eta_c^r$ and $0.87(k_B^2\theta)/h$ respectively (see Fig.~9 of Ref.~\cite{ya}). Thus, we see while maximum COP delivered is larger than that seen at the `Q' point of our  QR, the cooling power at that maximum COP is smaller. In Ref.~\cite{wang}, it has been shown that a magnon driven quantum dot refrigerator has COP $0.2\eta_c^r$ while again the cooling power seen $0.8(k_B^2\theta)/h$ is much smaller than that seen at the `Q' point of our system. Unfortunately, Refs.~\cite{ya, wang} do not discuss the `Q' point for their refrigerators. Further, Refs.~\cite{ya, wang} consist of a three terminal system, which by design has an advantage over a two terminal system since in a three terminal system heat and electric currents can flow between separate terminals, so one can have better control over these parameters. Similarly, we propose that in a quantum refrigerator based on three terminal strained graphene system the performance can be increased further, which can be the subject of another manuscript. {Further, our device works in the linear transport regime. The discussion on the non-linear transport regime is beyond the scope of this work. However, in a future work this aspect can be studied as whether the performance of the refrigerator increases more or not as compared to the linear transport regime.}      
\acknowledgments
This work was supported by funds from Science \& Engineering research Board, New Delhi, Govt. of India, Grant No. EMR/20l5/001836. 
   
\end{document}